\documentclass[twocolumn,secnumarabic,amssymb, nobibnotes, aps, prd, superscriptaddress
]{revtex4-1}

\usepackage[english]{babel}
\usepackage{graphicx}

\begin{document}

\title{Could short-lasting non-specific immunity explain seasonal spacing of epidemic diseases?}
\author{Gorm Gruner Jensen}
\thanks{G. G. J. and F. U. contributed equally to this work.}
\affiliation{Institute for Theoretical Physics, University of Bremen, D-28359 Bremen, Germany}
\author{Florian Uekermann}
\thanks{G. G. J. and F. U. contributed equally to this work.}
\author{Kim Sneppen}
\affiliation{Niels Bohr Institute, University of Copenhagen, Copenhagen 2100-DK, Denmark}
\author{Lone Simonsen}
\affiliation{Department of Science and Environment, Roskilde University, DK-4000 Roskilde,  Denmark}

\begin{abstract}
	Common respiratory viruses cause seasonal epidemics in sequential patterns. 
	The underlying mechanisms for this pattern have been debated for some time.
	For influenza, contenders include temperature, humidity, and vitamin D levels.
	While such seasonal drivers may be sufficient to explain midwinter peaking it is unclear if other respiratory viruses peaking in spring, summer, or fall have individual environmental drivers.
	Here we present a dynamic model of interacting diseases, in an effort to explain observed seasonal patterns without requiring multiple seasonal drivers.
	Our model extends the classical SIRS-model to include multiple diseases that interact via a short-lasting, non-specific immunity component.
	This temporal protection is triggered by exposure to any epidemic disease and lasts only a couple of weeks.
	We show that the inhibiting disease-interaction allows recurrent epidemic behaviour without any seasonal driving.
	In the presence of a single seasonal driver our model parsimoniously predicts epidemic patterns such as sequentially occurring epidemic diseases.
	We also present a two-disease simulation reproducing multiple features observed in time series of parainfluenza strains PIV-3 and biennial PIV-1 epidemic patterns.
	Complex epidemic patterns of seasonal diseases may be explained by non-specific innate or cloub cell (T-cell) mediated immune responses that result in interaction between unrelated respiratory epidemic diseases.
\end{abstract}

\maketitle


\section*{Keywords}
Seasonal epidemics; Seasonal ordering; Non-specific immunity; Disease interaction; SIRS model;




\section{Introduction}
Mysteriously, each year respiratory viruses cause epidemics that peak in predictable months during spring, summer, fall and winter 
\cite{hope1987new,nelson1999melatonin,smith2015update,fry2006seasonal,laurichesse1999epidemiological,counihan2001human}. 
The fact that the Northern and Southern hemisphere displays mirror image patterns of timing suggest that epidemic timing is determined by seasonal environmental driver(s). 
This has been known probably since the beginning of virology in the 1930s, but the phenomenon is still in search of an explanation.
Various authors have suggested that winter-seasonality in for example influenza is explained by a seasonal driver such as cold weather, low humidity or low vitamin D levels \cite{hope1987new,nelson1999melatonin,maes1994seasonal,boctor1989seasonal,dowell2001seasonal,cannell2006epidemic}. 
However, in order for such mechanisms to explain spring, summer, and autumn seasonality one would need to imagine a different seasonal driver for each season -- perhaps for each virus -- not an appealing or parsimonious hypothesis. 
Rather, the repeated annual patterns suggest that somehow epidemic diseases can repel each other and await their turn. 
Such a mechanism of “epidemic interference” has been proposed by Rohani
\cite{rohani1998population,rohani2003ecological,huang2005dynamical}
who demonstrated that by introducing short-lasting, non-specific immunity into classical dynamic childhood disease models one can find parameters for which one epidemic would exclude another. 
Yet most mathematical models of disease-disease interactions published to this date explore the effects of Ab-mediated specific, life long cross immunity
\cite{bhattacharyya2015cross,kamo2002effect,vasco2007tracking,sanz2014dynamics}.
When short-lasting immune mechanisms are considered it is typically in a single disease context to explain phenomena like multi-wave diseases
\cite{Camacho3635,saenz2010dynamics,pawelek2012modeling,antia1994model,cao2015innate} 
or to provide an evolutionary bottleneck for influenza evolution
\cite{ferguson2003ecological}.
Thus, models have yet to explored the possibilities of inter-epidemic dynamics in multiple-pathogen models that include terms of short-lasting non-specific immunity. 
Here we address this intriguing question anew, proposing a dynamic disease model form that can theoretically address this type of inter-species interaction. 
Our model work is motivated by the ever increasing evidence for short-lasting acquired non-specific immunity
\cite{cao2015innate,muraille2015unspecific,netea2011trained,netea2013training,hamilton2016club}.
Prominently, the mechanism has been directly demonstrated in controlled laboratory experiments with influenza A infections that prevented subsequent infection with influenza B for 6 to 12 weeks in mice
\cite{hamilton2016club}.

We explore theoretically whether it is possible to explain the phenomenon of recurrent multiple epidemics with predictable (and different) seasonality by introducing short-term non-specific immunity into the classical type of SIRS modelling and choosing parameter values consistent with known mechanisms such as innate immunity. 
This way we allow for individuals infected with a disease to be temporarily unavailable for infection with a second disease, and then re-enter the susceptible pool for the second disease after a few weeks. 
To test the performance of our model against observed epidemic patterns, we apply the model form to real-life epidemiological time series data of PIV-1 and PIV-3 in the US.
We find that our model can reproduce the observed “signature” patterns of seasonal diseases assuming only a single environmental driver and one type of short-lasting non-specific immunity.

Before introducing the model in detail we emphasise that there are multiple biological motivations for short non-specific immunity. 
While it may represent a universally active innate immunity
\cite{muraille2015unspecific}, 
any other less specific short-term immune defences that affect a class of diseases
\cite{dave1994viral,hamilton2016club} 
are valid motivations as well. 
We will not explicitly distinguish between these motivations in the following, as our model is general enough to represent any of them. 
As a result, it must be decided on an individual basis whether sub-types or strains of viruses should be considered distinct diseases in the context of this model or not. 
The crucial criteria are that diseases should not confer significant long-term cross-immunity to each other and allow for reinfection. 
For example, in the case of influenza A, we would consider the different sub-types (H3N2,H1N1) to be distinct diseases, since infection with one does not typically confer long-term immunity against another. 
The strains of a sub-type however would be considered the same disease here, to allow reinfection every couple of years. 
Under that representation influenza "waning" specific immunity become an abstraction of antigenic drift and cluster transitions.
\cite{plotkin2002hemagglutinin} 
With respect to PIV (parainfluenza virus) and RSV (respiratory syncytial virus) on the other hand, the waning immunity represents an actual loss of effective specific immunity in the host.
\cite{chanock1963myxoviruses,hall1991immunity} 

\section{Model and Methods}
\begin{figure}
	\includegraphics[width=\linewidth]{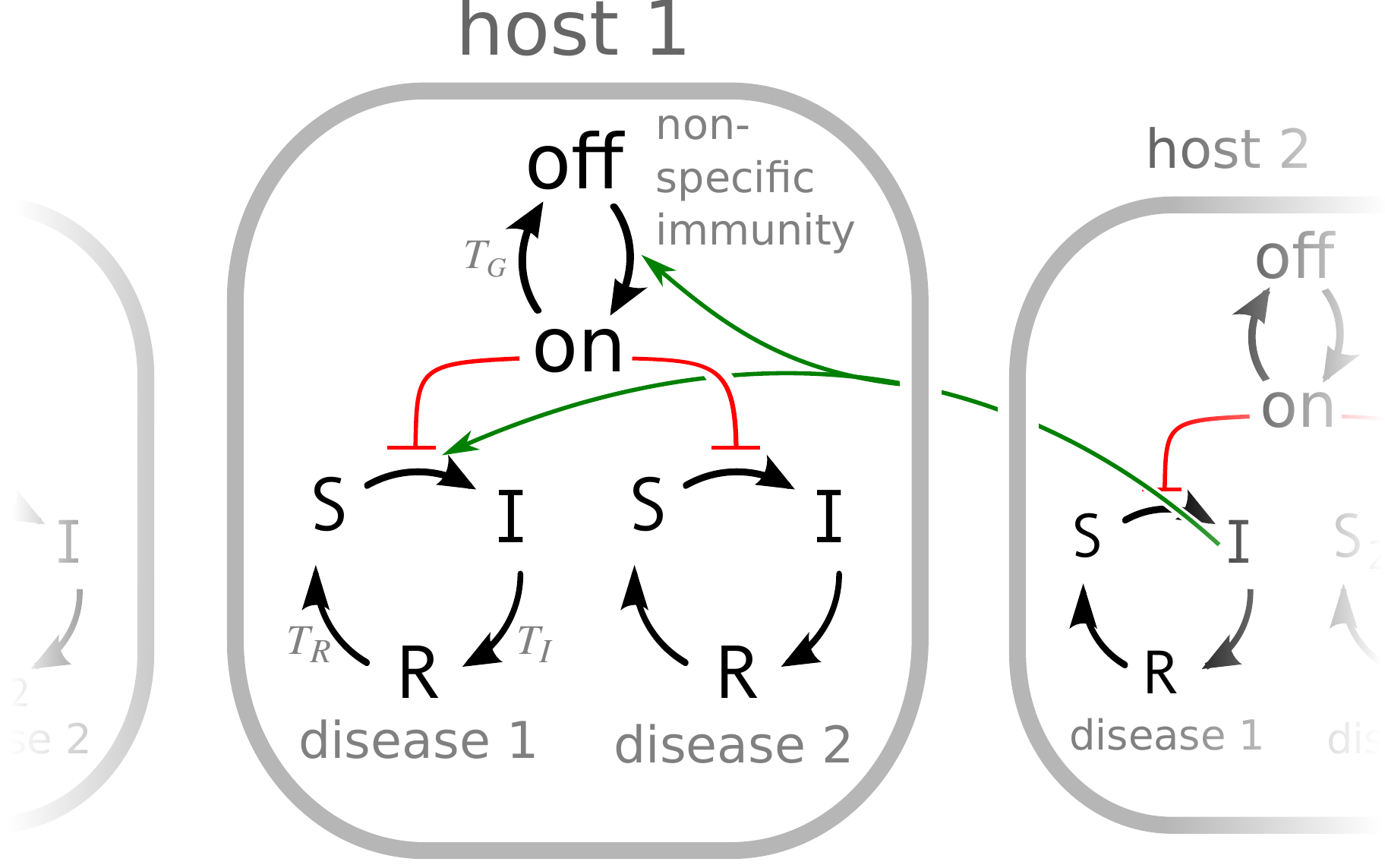}
	\caption{\label{fig:ModelDescription}
		Model from the perspective of individual hosts. Each host is in either the $I$, $R$, or $S$ state with respect to each disease, and  either has non-specific immunity ($G_{on}$) or not ($G_{off}$).
		The green arrows represent exposure of host 1 to disease 1 due to contact with the infected host 2.
		Infections ($S\rightarrow I$) require exposure to the disease, but are prevented if the the host has non-specific immunity $G_{on}$ (as indicated by the red links).}
\end{figure}

Our model is based on the traditional $SIRS$-model\cite{anderson1992infectious} with a homogeneous, fully mixed host population.
We extend this model to multiple diseases and add a new feature, the effect of a non-specific immune protection, which is short-lived.
We think of this additional kind of immunity as a way of modelling either an increased efficiency of the innate immune system observed by \cite{hamilton2016club} or as short-lived cross-immunisation between similar, but not identical, diseases.
All events, like contact between hosts or recovery from a disease, are modelled on the basis of transition rates.
In agent-based simulations with host-populations of finite size this leads to stochastic dynamics, while the behaviour for very large host-populations can be approximated using differential equations.
We will describe the model from the perspective of individual hosts, which is shown schematically in figure \ref{fig:ModelDescription}.
\paragraph*{SIRS:}
The traditional $SIRS$-model can be summarised as follows:
Each host can be {\bf S}usceptible, {\bf I}nfectious, or {\bf R}emoved.
The transmission of diseases is modelled using a contact rate.
If the contacted host is susceptible, it becomes infectious ($S \rightarrow I$).
During one period of being infectious, a host will on average contact $R_0$ randomly chosen hosts. 
This is commonly referred to as the basic reproduction number.
The average time a hosts stays infectious until entering the removed state ($I\rightarrow R$) is $T_I$.
Similarly, a removed host becomes susceptible again ($R\rightarrow S$) after an average time of $T_R$.

\paragraph*{Multiple diseases:}
To describe multiple diseases spreading in the same host population, we track whether the host is {\bf S}usceptible, {\bf I}nfectious, or {\bf R}emoved with respect to each disease independently (see disease $1$ and $2$ in Fig. \ref{fig:ModelDescription}).
The $R$ state therefore represents a disease specific immunity.

We then add a non-specific immunity, which is either active ($G_{on}$), or inactive ($G_{off}$).
If it is active, the host is protected against infection with any disease, even if it is in the $S$ state with respect to some of them.
The non-specific immunity is activated ($G_{off}\rightarrow G_{on}$) every time a host comes into contact with any of the diseases, also if infection is prevented by specific immunity (host is in $R$ state with respect to the disease).
This assumption allows maximal interaction between diseases.
The average time the non-specific immunity stays active is $T_G$ ($G$ for {\bf G}eneral Immunity).

Since the assumption of short-term but $100\%$ effective non-specific immunity is not necessarily realistic, we conducted simulations with non-specific immunity that is twice as long, but not fully protective ($50\%$).
We achieved both qualitatively and quantitatively similar results (data not shown).

\paragraph*{Model parameters:}
The model has 4 parameters, three of which are independent.
Throughout the paper we keep as many parameters as possible fixed to the same value.
\begin{description}
\item[$R_0$] Basic reproduction number. Mean number of transmission attempts over the course of one infection. Expected number of secondary cases in a completely susceptible population. $R_0=2.5$ in case studies without seasonal variation. Case studies with a seasonal driver use $R_0(t)=2+c\cdot\sin(2\pi t/\tau)$, with $c=0.1$ or $c=0.2$ and $\tau = 1$year.
\item[$T_I$] Mean duration of infectiousness. Since scaling all durations $T_X$ by a common factor has no dynamic effect, we set $T_I=1$ without loss of generality. For common respiratory diseases, $T_I=1$ represents $\approx 6$ days. All simulations are conducted with $T_I=1$, without exception. All other durations are specified in multiples of $T_I$.
\item[$T_R$] Mean duration of disease specific immunity. Unless stated otherwise $T_R=100$.
\item[$T_G$] Mean duration of short non-specific immunity $T_G\leq6$. See individual figure for specific values.
\end{description}
$R_0$, $T_I$ and $T_R$ can be disease specific parameters. Unless stated otherwise, they are chosen to be identical for all diseases.

The mean duration of specific immunity is rather short in most simulations (few years). This reflects the fact that immunity is often incomplete in viruses such as PIV\cite{chanock1963myxoviruses} and RSV\cite{hall1991immunity}, which allows for frequent reinfection. In influenza on the other hand, the effective duration of specific immunity is shortened by the antigenic evolution of influenza\cite{plotkin2002hemagglutinin}.

\paragraph*{Simulation details:}
All simulations discussed in this paper were initialised with an almost exclusively susceptible population.
Stochastic agent-based simulations with a finite number of hosts were performed using a Gillespie algorithm.
The population consists of $500\,000$ susceptible hosts and a single 'exterior host', who is infectious with all diseases and never recovers.
The 'exterior host' represents external sources of infections, which has the side-effect of preventing stochastic extinction.

Similarly, the deterministic simulations contain an 'external reservoir' of always infectious hosts which is $500\,000$ times smaller than the simulated population.
The external reservoir dramatically reduces transient time, i.e. the time before a stable pattern emerges.
In simulations with a seasonal driver the external reservoir is $200\,000$ times smaller than the simulated population.
This increase of external infection severely reduces the transient time before the epidemic limit-cycle is reached.

The deterministic rate equations are solved using the Euler method with a time-step of $0.001 T_I$.
To break the symmetry between diseases with identical dynamical parameters, we initialise the deterministic simulations with a one thousandth of the population being infectious with disease one, and two thousandths of the population being infectious with disease two, etc.

We have not systematically explored the robustness of the dynamics to initial conditions.
However, we also have not observed sensitivity to the choice of initial-conditions, except for a small parameter region of bi-stability marked in Fig.~\ref{fig:Phenomenon}c.

\section{Results}
First, we investigate the simplest case where interactions can be observed: Two diseases with identical dynamical parameters, without changing environmental factors.
Fig. \ref{fig:Phenomenon}a\&b shows two case studies with different durations of average of non-specific immunity ($T_G$), for which we observe qualitatively different behaviour.
In the case of very short non-specific immunity ($T_G = 0.1$), both diseases approach an endemic equilibrium state as shown in Fig. \ref{fig:Phenomenon}a.
This resembles the well known behaviour without non-specific immunity (traditional $SIRS$).
The other case, shown in Fig. \ref{fig:Phenomenon}b, considers a longer duration of non-specific immunity ($T_G=4$, corresponding to about 24 days for a disease with a 6-day infectious period).
Here we observe a repeated pattern of distinct epidemics, alternating between the two diseases.
The agent-based and deterministic (infinite population size) simulations exhibit the same qualitative behaviour.
These examples show that introduction of non-specific immunity can lead to recurrent alternating epidemics, a feature that unmodified deterministic $SIRS$ models do not exhibit\cite{korobeinikov2002lyapunov}.
These case studies show that the non-specific immunity needs to exceed a certain average duration to maintain recurrent epidemics.

In Fig. \ref{fig:Phenomenon}c we further investigate the phenomenon by measuring the amplitude of the recurring epidemics for specific choices of $R_0$ and duration of non-specific immunity $T_G$.
Since realistic durations for the non-specific immunity are short ($2-4$ weeks), we only show the behaviour for $T_G<15$, which corresponds to up to $3$ months, if the infectious period is $6$ days long.
In this range, recurrent epidemics are observed for most reasonable values of $R_0$ and $T_G$.
More specifically, recurrent epidemics are only absent if the interaction between the two diseases is very weak (short non-specific immunity time) or $R_0$ is close to $1$. 

The time series in Fig. \ref{fig:Phenomenon}a\&b show the transient behaviour of simulations.
Much longer simulations, such as those underlying Fig. \ref{fig:Phenomenon}c show that the behaviour indeed converges to stable periodic attractors.

\subsection{Recurring alternating epidemics}
\begin{figure}
	\includegraphics[width=\linewidth]{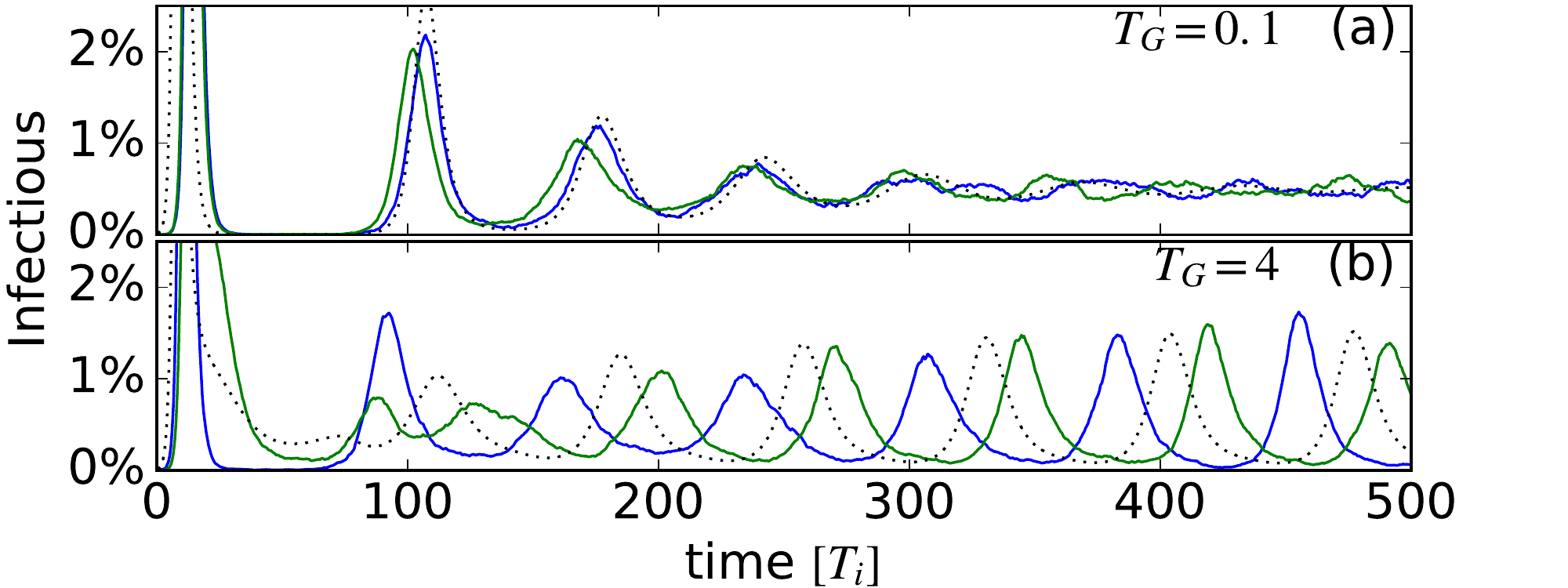}
	\includegraphics[width=\linewidth]{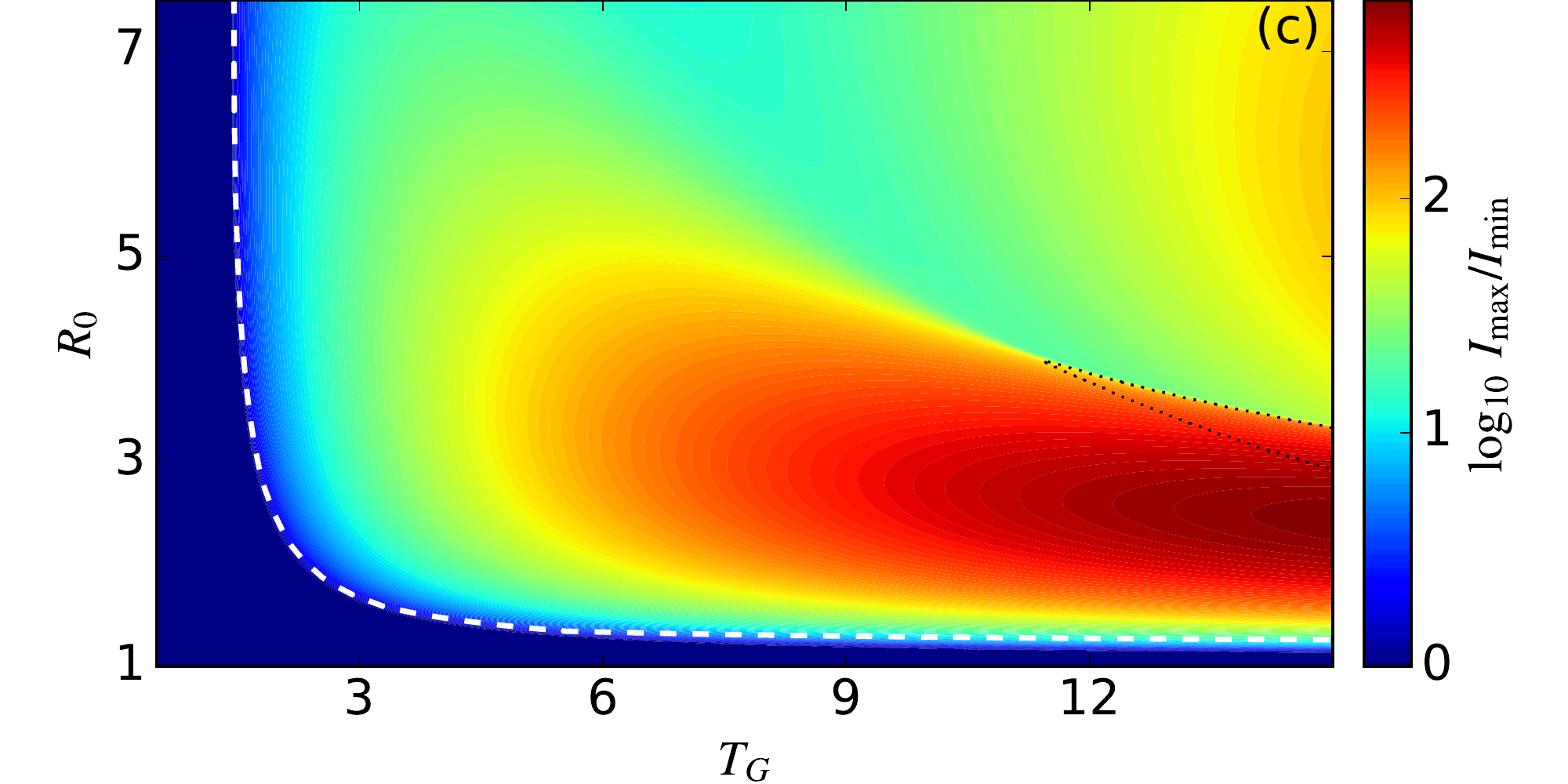}
	\caption{\label{fig:Phenomenon}
		Dynamics with two diseases with mean specific immunity duration of $T_R = 100T_I$. {\bf(a\&b)}: Case studies with $R_0 = 2.5$. Continuous lines show stochastic agent-based simulations with $500000$ hosts. Dotted lines show one of the diseases in deterministic simulations. {\bf(a)}: Short non-specific immunity $T_G=0.1T_I$, leads to endemic equilibrium state. {\bf(b)}: Longer non-specific immunity $T_G=4T_I$, leads to recurrent alternating epidemics. {\bf(c)}: Spikiness of epidemic behaviour. For each data point the deterministic simulation of two diseases is run for a period of $11000 T_I$. The maximum and minimum of infectious hosts ($I_{max}$, $I_{min}$) during the last $1000 T_I$ is measured and $\log_{10}\frac{I_{max}}{I_{min}}$ is plotted. The white dashed line was obtained by linear stability-analysis and marks the bifurcation line between stable endemic fixed point and epidemic limit-cycles. The black dotted line encloses an area with two stable epidemic limit-cycles (the plotted color corresponds to the one with larger amplitude).}
\end{figure}
For a given parameter set, we can numerically find the endemic fixed point and then perform a linear stability analysis.
We have done this on a fine grid over the parameter spaces shown in Fig. \ref{fig:Phenomenon}c \&  \ref{fig:Robustness}c and found supercritical Hopf-bifurcations, where the stable (endemic) fixed-point becomes unstable and emits a stable (epidemic) limit-cycle.
The bifurcation lines are shown as dashed white lines in Fig. \ref{fig:Phenomenon}C \& \ref{fig:Robustness}C.
The amplitudes measured in deterministic simulations match well with the calculated bifurcation lines.

\subsection{Diseases with different dynamical parameters}
\begin{figure}
	\includegraphics[width=\linewidth]{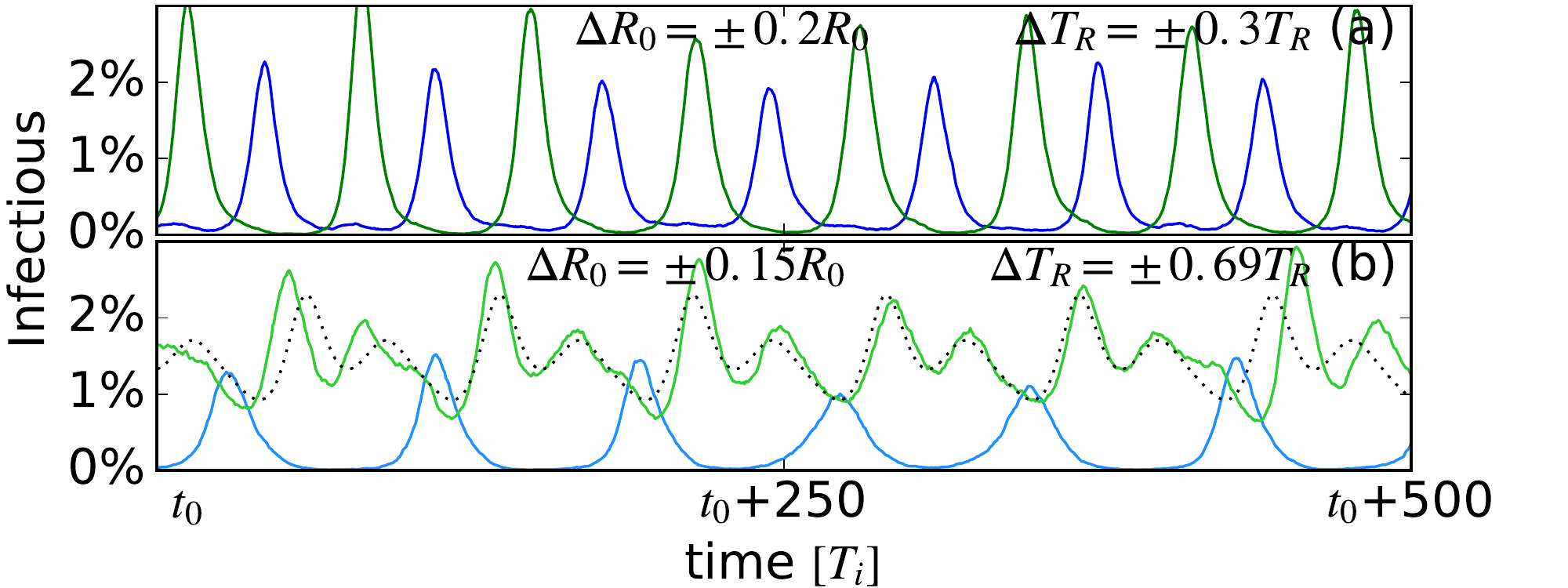}
	\includegraphics[width=\linewidth]{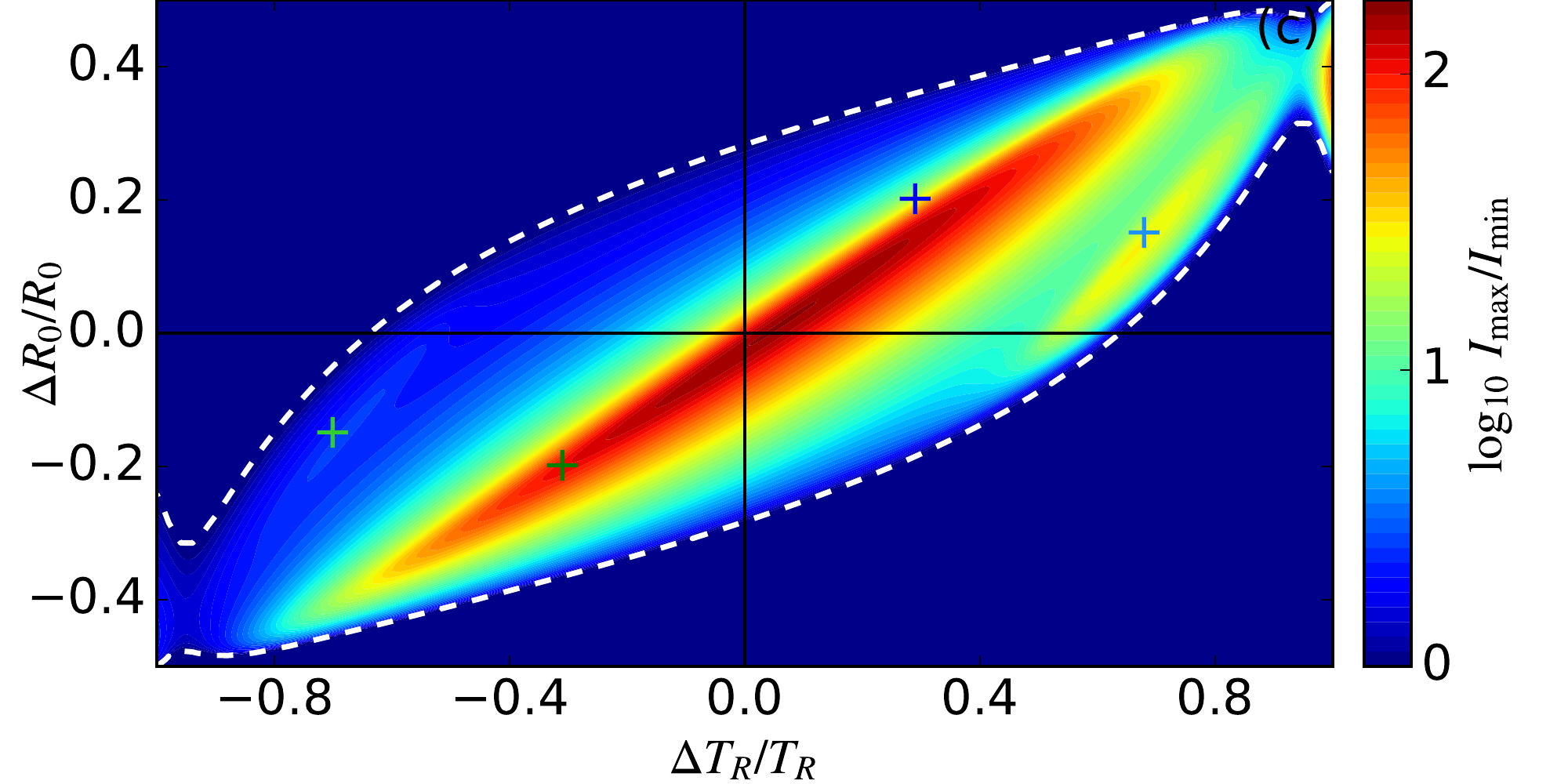}
	\caption{\label{fig:Robustness}
		Dynamics with two diseases with different dynamical parameters. The transmissibility $R_0$ and specific immunity duration $T_R$ differ between diseases by $\pm\Delta R_0$ and $\pm\Delta T_0$ from their respective mean $R_0 = 2.5$ and $T_R = 100T_I$. $T_G=6T_I$ remain unchanged, and $T_I$ is the same for both diseases.
		{\bf(a\&b)}: Case studies. Stochastic simulations are shown in solid lines. Dashed line shows the deterministic results for one of the diseases in b. $\Delta T_0$ and $\Delta R_0$ are marked in c in the corresponding line color. {\bf(c)}: Spikiness of epidemics. See Fig. \ref{fig:Phenomenon} for details on measurement.}
\end{figure}
In the previous section we established the emergence of recurrent epidemics for two diseases for a wide range of dynamical parameters, which were set to be identical for both diseases.
A natural concern is, whether this behaviour is robust to variations of dynamical parameters between the two diseases.
Next we explore the dynamics of two diseases with differing parameters for transmissibility ($R_0$ changes, while $T_I=1$ remains unchanged) and duration of specific immunity ($T_R$).
We approach this question by varying the disease specific $R_{0,d}$ and the duration of specific immunity $T_{R,d}$ of the two disease in opposite directions:
\begin{eqnarray}
	R_{0,d=1} = R_0 + \Delta R_0  \qquad\qquad	R_{0,d=2} = R_0 - \Delta R_0\\ 
	T_{R,d=1} = T_R + \Delta T_R  \qquad\qquad	T_{R,d=2} = T_R - \Delta T_R 
\end{eqnarray}
We choose $R_0=2.5$ and $T_R=100T_I$ as base values and set the average duration of non-specific immunity to $T_G=6T_I$.
As in the previous section, we will first examine two case studies, which are shown in Fig. \ref{fig:Robustness}a\&b.
The exact choice of parameters is shown in Fig. \ref{fig:Robustness}c.
In the first case study $R_0$ and $T_R$ are both increased for one disease and decreased for the other, which leads to a dynamics of alternating recurrent epidemics, similar to the pattern observed for diseases with identical parameters.
In the second case study the two parameters are varied in opposite directions, $R_0$ is increased, while $T_R$ is decreased for one disease and vice-versa for the other.
The epidemic pattern still shows recurring epidemics, however, the two disease exhibit a different number of peaks per period and less pronounced peaks.

We then explore this phenomenon in more detail (figure~\ref{fig:Robustness}c).
In general larger $R_0$ increases the frequency of recurrent epidemics in our model, while larger $T_R$ reduce the frequency.
The diagonal region with high amplitude (red) in the centre of Fig. \ref{fig:Robustness}c signifies parameter variations, with little change in peak frequency, where the differences in $R_0$ and $T_R$ compensate each other.
This produces alternating recurrent epidemics as seen in \ref{fig:Robustness}a.
Parallel to this central region are two secondary areas with high amplitudes, which represent parameter sets, where one peaks twice as often as the other, as seen in the second case study.
If the periods don't match, the sustained alternating epidemics are replaced by a more varied temporal pattern with lower amplitudes.

We find that the epidemic behaviour is robust to moderate differences in both $R_0$ and $T_R$, especially when the variations are such that the natural periods of recurrence of the two diseases are similar or approximately multiples of each other.

\subsection{Multiple Diseases}
\begin{figure}
	\includegraphics[width=\linewidth]{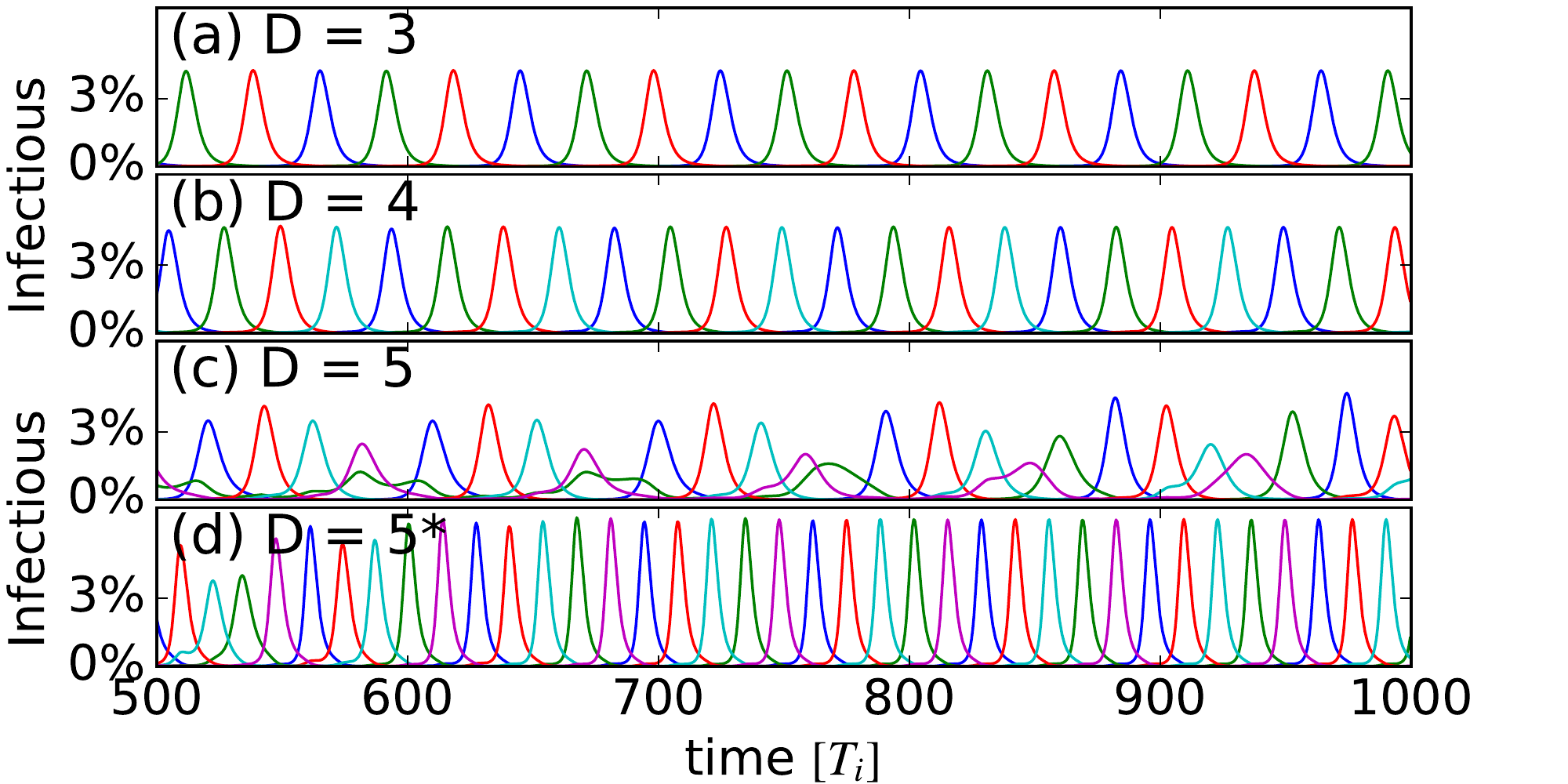}
	\caption{\label{fig:NDiseases}
		More than $2$ diseases in deterministic simulations with $T_G=6T_I$. The basic reproduction number is $R_0=2.5$ except for panel (d), where $R_0=5$. The mean specific immunity time is $T_R = 100T_I$.}
\end{figure}
After exploring the case of two diseases in detail, we will briefly show examples with more than two diseases. 
The model is defined for an arbitrary number of diseases, so it's straight forward to extend our investigations.
The number of possible host-states however, grows exponentially with the number of diseases, which quickly makes the simulations unfeasible in the deterministic limit of infinitely many agents.
'Individual host'-based simulations could be a way around this issue, but it would not be possible to study the model without noise.
Deterministic simulations for up to $5$ diseases are shown in Fig \ref{fig:NDiseases}.
With increasing number of diseases the temporal pattern becomes more varied.
In general only systems with a limited number of diseases can sustain alternating recurrent epidemics.

We observe that larger values of both $R_0$ and $T_R$ allow for narrower epidemic spikes, which increases the number of supported diseases without breaking the pattern of alternating epidemics.
Narrower epidemic spikes are necessary to sustain many alternating recurrent epidemics, because the  non-specific immunity suppresses simultaneous outbreaks of different diseases, thus limiting the fraction of time available to the individual disease.

\subsection{Seasonality}
\begin{figure}
	\includegraphics[width=\linewidth]{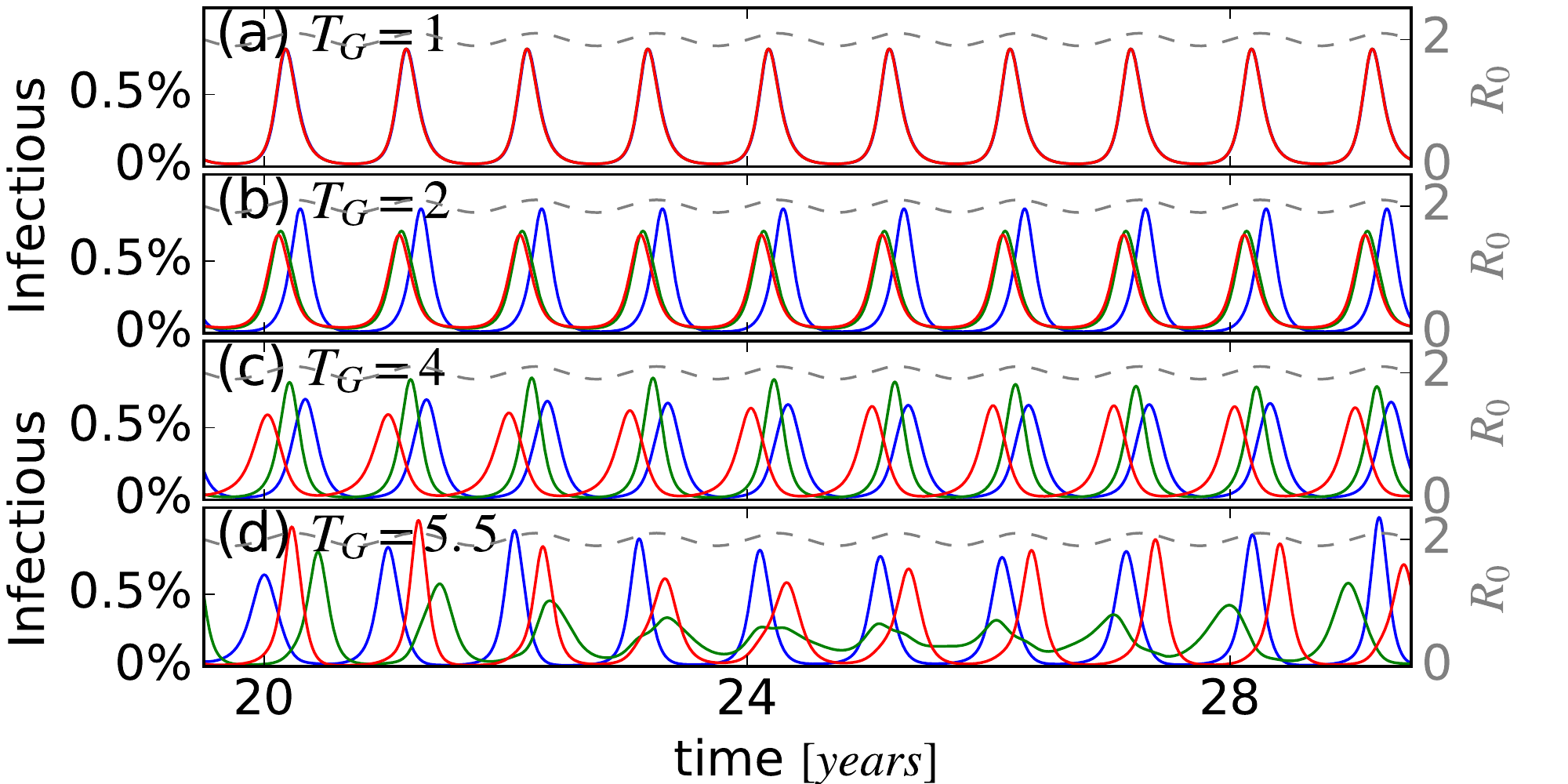}
	\caption{\label{fig:seasonality}
		Deterministic simulations with three disease-systems with different strength of non-specific immunity $T_G$. All dynamical parameters except for $T_G$ are equal in all panels and for all diseases: 
		$R_0(t) = 2 + 0.1\cdot\sin (2\pi t/120) $, 
		$T_R = 250T_I$.
		{\bf (a)} With weak interaction (low values of $T_G$), the seasonal driver dominates the dynamics. 
		{\bf (b\&c)} With intermediate interaction-strength the seasonal driver controls the periodicity, but the epidemic peaks are separated in time. 
		{\bf (d)} Strong interaction prevents completion of distinct epidemics during every year. In combination with the seasonal driver this leads to a loss of the annual periodicity.
	}
\end{figure}
A common method for incorporating seasonal factors in \textit{SIRS}-models is by making $R_0$ time-dependent.
Consequently, this section will discuss an extension of our model with an annually varying basic reproduction number $R_0(t)=2 + 0.1\cdot\sin (2\pi t/\tau)$, where $\tau$ specifies the length of a year in multiples of $T_I$.

Fig. \ref{fig:seasonality} shows different 3-disease patterns produced by our model for different choices of the duration of non-specific immunity $T_G$.
In this context we can think of $T_G$ as a control-parameter for the strength of interaction between the diseases.
It is worth noting that a similar set of patterns can be found by varying $T_R$, or the mean-value or amplitude of $R_0(t)$.
We use the same dynamical parameters for all three diseases and subject them to the same seasonal driving via $R_0(t)$.
Similar to the multitude of observed epidemic patterns in nature, our model produces a variety of dynamics, based on the exact parameter choices.
In the following we will relate some patterns observed in nature to the different panels in Fig. \ref{fig:seasonality}:

In the case of concurrently spreading influenza A sub-types (H3N2 and H1N1) whose concurrently dominant strains have different antigenic profiles, we see that the dynamics is dominated by the seasonal driver, with few differences between disease variants.
Unsurprisingly, this is reproduced in panel a, where the non-specific immunity is short, which implies weak interaction.

Some types of PIV on the other hand, tend to peak at different times than others. PIV-3 typically peaks in early summer, while PIV-1, 2 and 4 are more common in the fall.
Panel B shows a similar pattern between diseases with medium duration of non-specific immunity. 
One disease peaks separately from the other two, which remain synchronised.

Panel C shows the behaviour with non-specific immunity that lasts long enough to spread the peaks of all three diseases over more than half a year, even though they have the same dynamical parameters and depend on the same seasonal driver.
Finally, in panel D, the interaction between diseases is so strong that the epidemic behaviour does not converge to a pattern with a one year periodicity.

\subsection{Modelling observed PIV-1\&3 seasonal pattern}
\begin{figure*}
	\includegraphics[width=\textwidth]{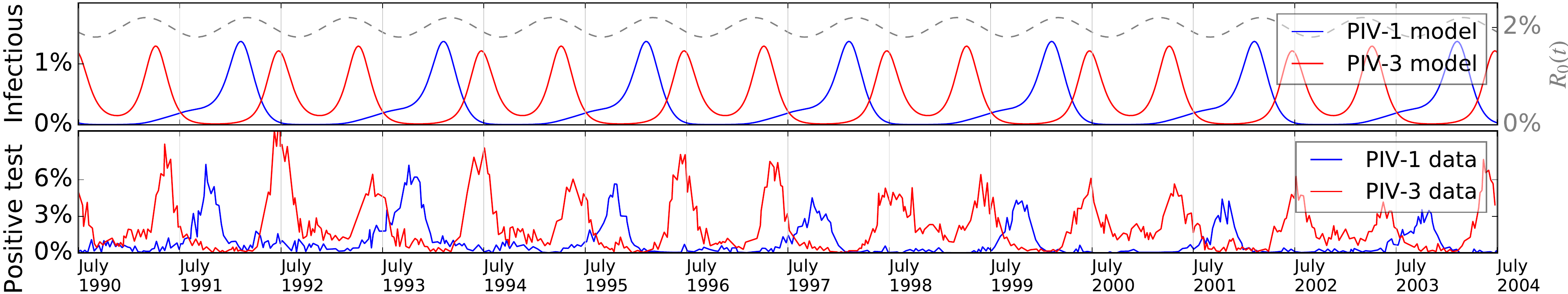}
	\caption{\label{fig:PIV}
		Upper panel: Seasonal variant of the deterministic model with two diseases, resembling PIV-1 and PIV-3. The basic reproduction number is varied seasonally using a sine function $R_0(t)=2+0.2\cdot\sin(2\pi t/(60 T_I))$. The average immunity time is $T_G=4T_I$ and the specific immunity times are $T_{R,PIV1}=170T_I$ and $T_{R,PIV1}=120T_I$. Rescaled to a year, this implies an infectious period of $\approx6$ days, on average $\approx24$ days of non-specific immunity, and specific immunity that lasts on average $2$ years (PIV-3) and $\approx2.8$ years (PIV-1). The external reservoir of constantly infectious hosts is $1 / 200\,000$ of the population size. The annual maxima of $R_0$ are aligned with March 1st in the lower panel.
		Lower panel: Percentage of positive tests for PIV-1 and PIV-3 serotypes in the US reported to NREVSS from July 1990 to June 2004. This data was extracted from {\it Fry et al.}\cite{fry2006seasonal}. Due to the low image quality of the original, the estimated extraction uncertainty is $(\pm10\text{days},\pm0.2\%)$.}
\end{figure*}
For PIV-1 and PIV-3 a particularly interesting pattern is typically observed.
The activity of PIV-3 between epidemic peaks is consistently lower every second year, which coincides with the biennial outbreaks of PIV-1 \cite{fry2006seasonal,laurichesse1999epidemiological}.
It has been suggested that this correlation is caused by an interaction between the two diseases\cite{fry2006seasonal}.

We found that our model can reproduce multiple features of the described dynamics, with a seasonally varying $R_0$ and a longer specific immunity-time for PIV-1 than for PIV-3 (see Fig. \ref{fig:PIV} for detailed information on parameters).
The direct comparison between simulated and observed prevalence in Fig. \ref{fig:PIV} shows that the model reproduces many features of the observed dynamics, such as: (1) PIV-3 peaks annually, while PIV-1 peaks every second year. (2) In years with a PIV-1 epidemics, the valley of PIV-3 activity is significantly deeper. (3) The next PIV-3 epidemics is delayed by $\approx 3$ months in years with a PIV-1 epidemic. (4) PIV-1 epidemics ascend for many months and end abruptly, while the individual PIV-3 peaks are more symmetrical.
If the simulation is executed with zero -- or very low levels of -- non-specific immunity, but otherwise identical parameters, the system converges to a limit-cycle in which both diseases have epidemic outbreaks every year. 

\section{Discussion}
Many respiratory viral diseases cause seasonal epidemics in predictable months.
It is well known that epidemics of different diseases spike at different times of the year, but it is not clear why.
Typical discussions of epidemic patterns point to a variety of seasonal drivers\cite{hope1987new,nelson1999melatonin,maes1994seasonal,boctor1989seasonal,dowell2001seasonal,cannell2006epidemic}.
It is not clear, however, why these would have a different influence on otherwise similar respiratory pathogens causing these epidemics.
Here we have demonstrated that a simple SIRS-model with a single seasonal driver can predict different seasonal patterns even for diseases with identical dynamical parameters (figure~\ref{fig:seasonality}), provided that these diseases interact through a short-lasting non-specific immune response in their hosts.

At an organism level, innate immunity, or cross-immunity between specific viral strains, may have many different causes\cite{cao2015innate,muraille2015unspecific,netea2011trained,netea2013training,hamilton2016club}.
Our non-specific immunity mechanism could be interpreted as any of these or even a combination.
It is, however, worth noting that the effect should be short-lasting, i.e. much shorter than the effect of antibody mediated (pathogen specific) immunity, but also not too short.
In figure~\ref{fig:seasonality}a it is observed that if the average duration of the non-specific immunity is as short as the average infectious period, the interaction between diseases is too weak to overcome the effect of the environmental driver.
This seems to indicate, that the observed effect would not occur just because of behavioural changes caused by the sickness, such as staying home from work.

This paper is not the first to make a theoretical investigation of how non-specific immunity may change the epidemic patterns of diseases.
Some works have explored a similar concept using a coalescence, following the infection of one disease, in which hosts are neither infectious nor susceptible to any diseases\cite{rohani1998population, rohani2003ecological,huang2005dynamical}.
However, these models are mostly inspired by phenomena related to childhood diseases, which leads to some fundamental differences in the underlying single-disease dynamics.
In particular, the pathogen-specific immunity is assumed to last forever, making it necessary to explicitly keep track of birth and death processes.
In addition, these models have, to the best of our knowledge, not yet been used to directly address the question of seasonal ordering.

We have found that our model is able to predict recurrent epidemic oscillations, even in the absence of seasonal driving.
This is an interesting result in itself, as it is well known that epidemic limit cycles don't exist for any parameters in the classical SIRS model.
In the figures~\ref{fig:Phenomenon} and \ref{fig:Robustness}, we have investigated these recurrent epidemic patterns using a combination of local linear stability analysis and direct simulations.
The reason why non-specific immunity can break the stability of the endemic fixed point in our model is that one disease postpones the spread of others, thereby increasing the number of hosts without specific immunity to these temporarily suppressed diseases.
In turn, this causes larger epidemics once the non-specific immunity wears off, thus leading to epidemic peaks, as opposed to stable levels infectious hosts.
While these results are interesting, they are not particularly surprising since similar phenomena have been observed in related models with short term non-specific immunity\cite{rohani1998population, rohani2003ecological,huang2005dynamical}, as well as epidemic models with life long cross immunities\cite{bhattacharyya2015cross,kamo2002effect,vasco2007tracking,sanz2014dynamics}.

Finally we have shown an example of how our model is able to reproduce a number of qualitative features in the complex epidemic pattern of PIV-1 and PIV-3 observed in the period between 1990 and 2004 in the United States.
This strengthens the hypothesis proposed by Fry et. al\cite{fry2006seasonal}, that this pattern is caused by interference between the epidemics of the two diseases.
With this result in mind, we suggest that disease-disease interactions mediated by non-specific immunity (innate immunity) may have a larger effect epidemic patterns than previously suspected.

\section*{Data, code and materials}
The model is completely reproducible from the description in the main text, and all parameters are made available in the figure captions.

\section*{Competing interests}
We have no competing interests

\section*{Authors' contributions}
Gorm Gruner Jensen wrote most of the simulations, participated in developing the model; 
Florian Uekermann helped producing code, and extracted the time-series data shown in figure \ref{fig:PIV}, participated in developing the model;
Lone Simonsen conceived the idea of studying the seasonal ordering of infectious diseases; 
Kim Sneppen coordinated the collaboration, participated in developing the model; 
All authors participated in designing the study, draft the manuscript, and gave final approval for publication.

\section*{Funding}
This research has received funding from the European Research Council under the European Union's Seventh Framework
Programme (FP/2007 2013)/ERC Grant Agreement n. 740704.

\bibliographystyle{unsrtnat}
\bibliography{bibliography}



\end{document}